# Coverage measurement in model-based testing of web applications: Tool support and an industrial experience report

Vahid Garousi
Queen's University Belfast, UK
Testinium A.Ş., Türkiye
ProSys MMC, Azerbaijan
v.garousi@qub.ac.uk

Alper Buğra Keleş, Yunus Balaman,
Alper Mermer
Testinium A.Ş., Türkiye
{alper.keles, yunus.balaman, alper.mermer}@testinium.com

Zeynep Özdemir Güler
ING Türkiye, Türkiye
Zeynep.OzdemirGuler@ing.com.tr

**Abstract--**There are many widely used tools for measuring test-coverage and code-coverage. Test coverage is the ratio of requirements or other non-code artifacts covered by a test suite, while code-coverage is the ratio of source code covered by tests. Almost all coverage tools show a few certain subset of coverage values, and almost always either test-coverage or code-coverage measures. In a large-scale industrial web-application-testing setting, we were faced with the need to "integrate" several types of coverage data (including front-end and back-end code coverage with requirements coverage), and to see all of them "live" as large model-based test suites were running. By being unable to find any off-the-shelf toolset to address the above need, we have developed an open-source test coverage tool, specific for MBT, named *MBTCover*. In addition to code coverage, the tool measures and reports requirements and model coverage, "live" as a given MBT test suite is executing. In this paper, we present the features of the MBTCover tool and our experience from using it in multiple large test-automation projects in practice. Other software test engineers, who conduct web application testing and MBT, may find the tool useful in their projects.

## 1 INTRODUCTION

In the context of software testing, "coverage" is a metric that measures the amount / extent of testing performed by a test set (suite).

Coverage can be measured from both black-box and white-box testing perspectives. The former case is called test coverage, and is the ratio of requirements, models or even user-interface (UI) artifacts, of the System Under Test (SUT), exercised (covered) by a given test suite. The latter case is called code coverage and is the ratio of source code exercised by a given test suite.

Measuring coverage provides various benefits, e.g., it is "*a useful tool for finding untested parts of a codebase*" (martinfowler.com/bliki/Test Coverage.html).

A Google search for "test coverage" and "code coverage", as of December 2023, returned around 4.8 and 3.2 billion hits, respectively, indicating the wide popularity and usage of coverage analysis in industry.

There are hundreds or even thousands of code and test coverage tools, which are used in different test automation contexts and for different programming language. For example, a coverage tool named EclEmma (eclemma.org), as a plug-in of the Eclipse Integrated Development Environment (IDE), is widely used in JUnit testing.

Given the widespread adoption of web and mobile applications, automated testing of and measuring coverage in the context of these applications have become major trends in testing [1]. For testing web and mobile applications, both front-end (client-side) and also back-end (server-side) components of a given SUT should be systemically tested and the code/test coverage values shall be analyzed. There are various coverage tools for web and mobile applications. For example, a few of the many coverage tools for JavaScript (JS) front-end code are: Jest (jestjs.io), Istanbul (istanbul.js.org), and the DevTools (Developer tools) feature of Google Chrome (developer.chrome.com/docs/devtools/coverage).

For measuring back-end coverage (code running on server side), there are also various tools, depending on the server-side technology of a given SUT, e.g., JaCoCo (jacoco.org) for server applications developed in Java, and PVOC (github.com/krakjoe/pcov) for applications developed in PHP.

While all the above coverage tools are robust and popular for their purposes, in the context of various industrial automated model-based testing (MBT) projects in our context (see the previous experience paper in [2]), we were faced with the following test-coverage need: to gather and present *both* client-side and server-side code-coverage data, plus a number test coverage data (such as requirements coverage), and present them "live" as a given MBT test suite is executing.

Furthermore, when conducting automated MBT of web applications using MBT tools such as GraphWalker (graphwalker.github.io), connected with the Selenium browser automation framework (selenium.dev), there are no off-the-shelf coverage tool which would connect to the (integrate with) MBT tools such as GraphWalker seamlessly, i.e., without the need for further integration code development.

To address the above need, we decided to develop a test coverage tool, named *MBTCover* (coverage for MBT), which gathers and provides the following four types of coverage measurements, when a MBT suite is running using the popular MBT tool GraphWalker:

- Code coverage:

1. Code coverage of front-end (client-side) JavaScript code of the web application under test
2. Code overage of back-end (server-side) code (Java in our case context)
- Test coverage:
3. Requirements coverage, defined as the ratio of the use-case steps covered so far by the running MBT suite
4. Model coverage, e.g., the ratios of nodes and edges in the MBT test models (with activity-diagram formalism), covered so far in the MBT execution

In this paper, we present the features of the MBTCover tool and also our experience from using it in multiple large test-automation projects in practice. The rest of this paper is structured as follows. Background and related work are discussed in Section 2. MBTCover tool, its architecture and features are described in Section 3. We discuss in Section 4 our industrial experience in using the tool. Finally, Section 5 concludes the paper.

## 2 BACKGROUND AND RELATED WORK

### 2.1 Model-based testing (MBT), and MBT in practice

As of this writing in 2023, MBT has been around for more than 50 years. An IBM technical report [3], published in 1970, is often referred to as one of the first known reported applications of MBT. The modeling semantics used in that first paper was Cause-Effect Graphs, and a prototype tool, named TELDAP (TEst Library Design Automation Program), for generating test cases was presented. A very large number of papers and reports have been published in MBT since then, by following different approaches to MBT, e.g., from the standpoints of model semantics (UML models, BPMN or other model types), level of modeling abstractions, test execution modes (offline or online), and test selection criteria (model coverage, fault-based, etc.) [4-6]. However, many studies report that: "*most developers [still] don't view MBT as a mainstream [testing] approach*" [7].

While it seems that most of MBT literature have been studies which conducted in academic and lab settings, a subset of the literature are studies conducted in practice and industrial contexts. We review a few selected studies below.

An author with affiliation in both industry and academia reported his view of the state of the art and challenges of "industrial-strength" MBT [8]. The reported experience and opinions are based on a MBT tool named RT-Tester, developed by the author's team. The paper highlights the importance of selecting the right modelling "formalism" for the testing problem at hand, and the fact that development of models, properly, can prove to be a major hurdle for the success of MBT in practice. As a related factor, the required skills for test engineers developing test models are significantly higher than for test engineers writing conventional test procedures. Other key factors for successful industrial-scale application of MBT as reported in the paper were: tracing requirements to the model, and automated compilation of traceability data.

An experience report of introducing MBT in the context of a system named European Train Control System (ETCS), developed by a large European company, named *Thales,* was reported in [9]. The authors argued that MBT is not applicable "out-of-the-box", and application of MBT in a given environment (industrial context) requires specific adaptations. The selected test model formalism was UML/OCL. Certain toolchain-specific model revisions had to be made, e.g., timed triggers had to revised in the UML semantics (meta-model). The team used Borland Together for formalizing and concretizing system models. The last sentence of the paper was: "*it seems like the industry may already be aware of the possible benefits of MBT but fears the issues and costs of its integration*".

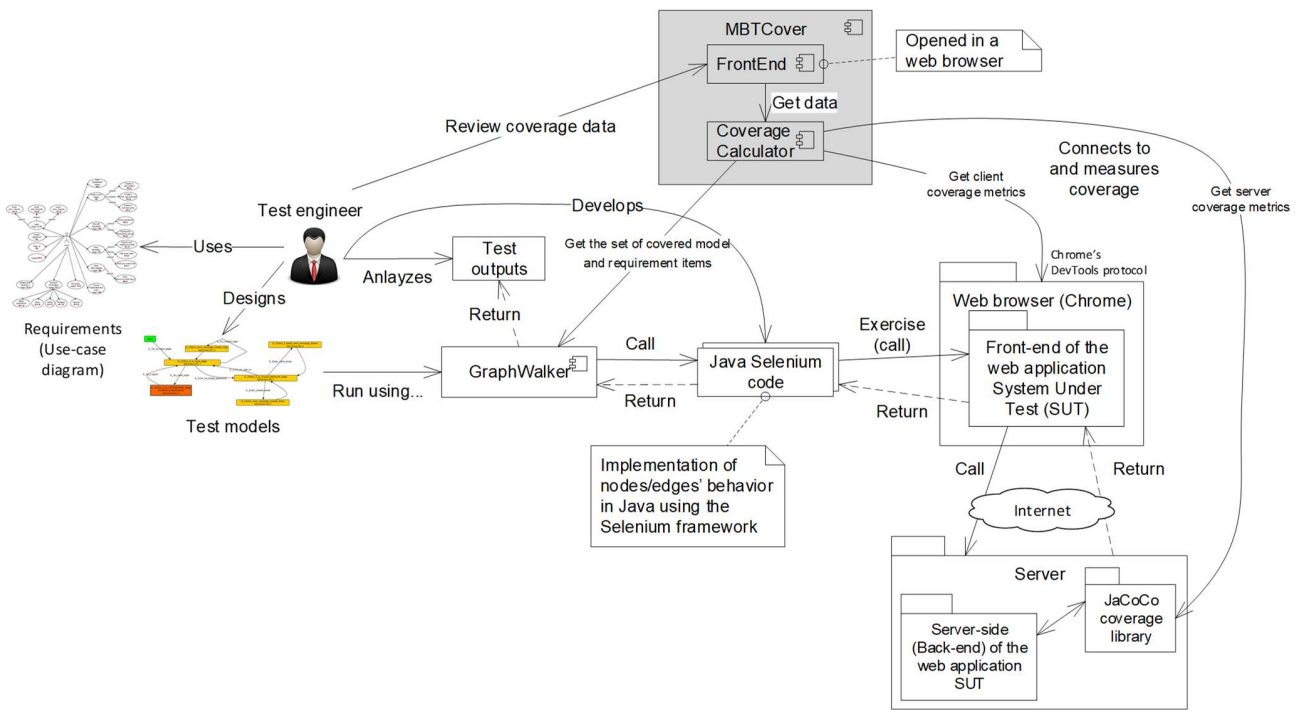

**Figure 1- Architecture and usage context of the MBTCover tool**

Microsoft has been one of the companies from which many MBT papers have been published, e.g., [10-12]. A 2003 paper [10] authored by a test architect at Microsoft reported the obstacles and opportunities for MBT in Microsoft. The author reported that: "*Model-based testing can provide a tremendous increase in testing capability, but modeling technology must be integrated into everyday software testing. Small-scale pilot projects, readily available tools and tester education have made the migration to test generation easier at Microsoft*". The author then reviewed how each of those characteristics affected the promotion of MBT at Microsoft. According to the paper [10], as of 2003, more than 600 of all 5,000 testers, working with Microsoft, were involved in some form of MBT.

Several papers from Microsoft have also presented their success story with MBT of documentation and quality assurance of client–server and server–server protocols of Microsoft Windows [11, 12]. A Microsoft MBT tool named SpecExplorer was used in those studies. The project was a large-scale undertaking in MBT: More than 25 000 pages of documentation for over 250 protocols had to be thoroughly verified to ensure that they are accurate, so that developers can implement protocols from the information they contain. Application of MBT reflected an investment of over 50 person-years. In addition, a substantial time investment was made in tool development, based on a continuous feedback loop from the test-suite development process into the SpecExplorer development team. According to statistical analysis, MBT resulted in a 42% productivity gain when compared with traditional test suites in a site where similar numbers of requirements were verified.

An interesting "voice of evidence" paper about MBT was published in IEEE Software in 2008 [7], which was based on systematic literature review (SLR). The authors argued that a rich body of experiences has not yet been published on all the SE techniques that researchers have proposed, including MBT. In fact, by some estimates, the techniques for which we do have substantial experience are few and far between. Thus, our current paper is a suitable evidence/experience paper aiming to address that gap. Based on their experience, the authors reported that: "*most developers [still] don't view MBT as a mainstream [testing] approach*" [7]. The study reported a "*serious lack of evidence*" in usefulness of different MBT approaches [13], and that many publications on MBT provide *only toy examples* without proper comparison with other approaches. The SLR divided the MBT studies into five categories: speculation, example, proof of concept, experience/industrial reports, and experimentation. UML-based MBT models were by far the most widely used formalisms. Furthermore, since applying MBT has non-trivial costs, the associated cost-benefits should be carefully analyzed when considering MBT, a topic referred to as "value-based" SE [14]. The study discussed this issue by stating: "*it's risky to choose an MBT approach without having a clear view about its complexity, cost, effort, and skill required to create [develop] the necessary models*" and that: "*Evidence on these topics could be a useful step in determining whether wider deployment of MBT approaches to different domains is worthwhile*".

### 2.2 Coverage measurement in testing web applications

There is (very) limited academic literature on coverage measurement in testing web applications, e.g., empirical studies. However, there are various sources in the grey literature on the topic, e.g., in the context of using the popular Selenium test automation framework[1]. In our literature review, we actually came across an online forum in which a practitioner was asking others, whether it is possible to collect front-end test coverage data when testing using Selenium[2]. This was additional motivation for the toolset that we have developed and present in this paper.

## 3 MBTCOVER TOOL

### 3.1 Usage context and architecture of the tool

We show in Figure 1 the usage context and software architecture of the tool. MBTCover supports test engineers in the context of MBT of web and mobile applications, by showing test coverage information.

MBT test models are designed by test engineers, in the form of activity diagrams showing the UI flow across different pages of a web application under test. For our large number of industrial MBT projects, from among a set of candidate tools, we systematically selected the MBT tool *GraphWalker* in our industrial context back in 2019, and have used it in several projects and previous papers, e.g., [2].

To use the MBT tool GraphWalker, to make MBT models directly executable, test engineers develop Selenium Java code to "implement" the action (behavior) of each node/edge in the models.

MBT models can then be executed, fully automated, using the MBT tool GraphWalker which uses the developed Selenium Java code to exercise (call) the front-end of the web application under test, and the front-end in turn communicates with the SUT's back-end. Test outputs are recorded, logged and returned to test engineers by the test tool GraphWalker.

To develop MBTCover, we selected a number of client-side and server-side coverage measurement technologies (details next) and integrated their outputs to show the results live visually, as a MBT suite is running. To get front-end JS coverage values at runtime, from among several alternative coverage tools, we selected the Chrome DevTools protocol, due to its stability, maturity and wide use in industry. To programmatically extract coverage data live from DevTools at runtime, we used a library called Puppeteer (pptr.dev) which provides an Application Programming Interface (API) to the Chrome browser.

---

[1] google.com/search?q=test+coverage+masurement+selenium

[2] quora.com/Is-it-possible-to-collect-front-end-test-coverage-statistics-with-Selenium

To get back-end (server-side) coverage live at runtime, and since the server-side language of almost all of our SUTs were Java, we have used the JaCoCo code coverage library (jacoco.org), due to its stability and maturity. We shall note that, if the reader intends to use MBTCover for other SUTs which have been developed in other programming languages (such as .Net), other server-side code coverage technologies can be easily integrated into MBTCover, instead of JaCoCo.

To better present and understand the MBTCover tool and how it works, we show a running example of a test model, designed and executed by the MBT tool, GraphWalker, and then MBTCover is used to measure coverage when running the MBT test suite.

### 3.2 An example MBT test suite

We show in Figure 2 one of the MBT models of a large MBT test suite for testing the *login* page of an industrial SUT named Testinium (testinium.io) [2]. Testinium is a web-based test management tool, which uses a number of test frameworks such as Selenium and Appium, and is one of the main products offered by the company (Testinium A.Ş.). the Testinium tool is in active use by 700+ test engineers of the company (Testinium A.Ş.) and also by many clients of Testinium A.Ş.

Let us note that this large MBT model suite includes 18 test models (each designed as a separate activity diagram), and contains in total 177 nodes and 260 edges. The entire MBT model is available open-source: github.com/vgarousi/MBTof Testinium.

In the last several years, in addition to MBT automated testing of the company's own products such as Testinium, we have completed more than 10 large MBT test projects for various client companies, e.g., MBT automated testing of a large bitcoin trading provider, with a very extensive web application, including more than 100 dynamic web pages, and two mobile applications for each mobile platform (Android and iOS). Some details of that MBT project can be found in: bit.ly/MBTofCOVIDandBanking. As stated above, in addition to web applications, our MBT projects also include MBT of mobile apps, e.g., one of recent consulting projects was MBT testing of the UK's three COVID contact-tracing apps. Details and test run videos can be found in github.com/vgarousi/MBT ofCOVIDapps.

As shown in Figure 2, the semantic of test models in the MBT tool GraphWalker is a form of UML activity diagrams showing the UI flow across different pages of the web application under test. In this MBT approach, as required by the MBT tool, test assertions (verification points) are placed in the nodes of activity diagram models (by implementing the suitable Selenium test-code); and web-page transitions (events, e.g., button clicks) are placed on the edges (again by implementing the suitable Selenium test-code).

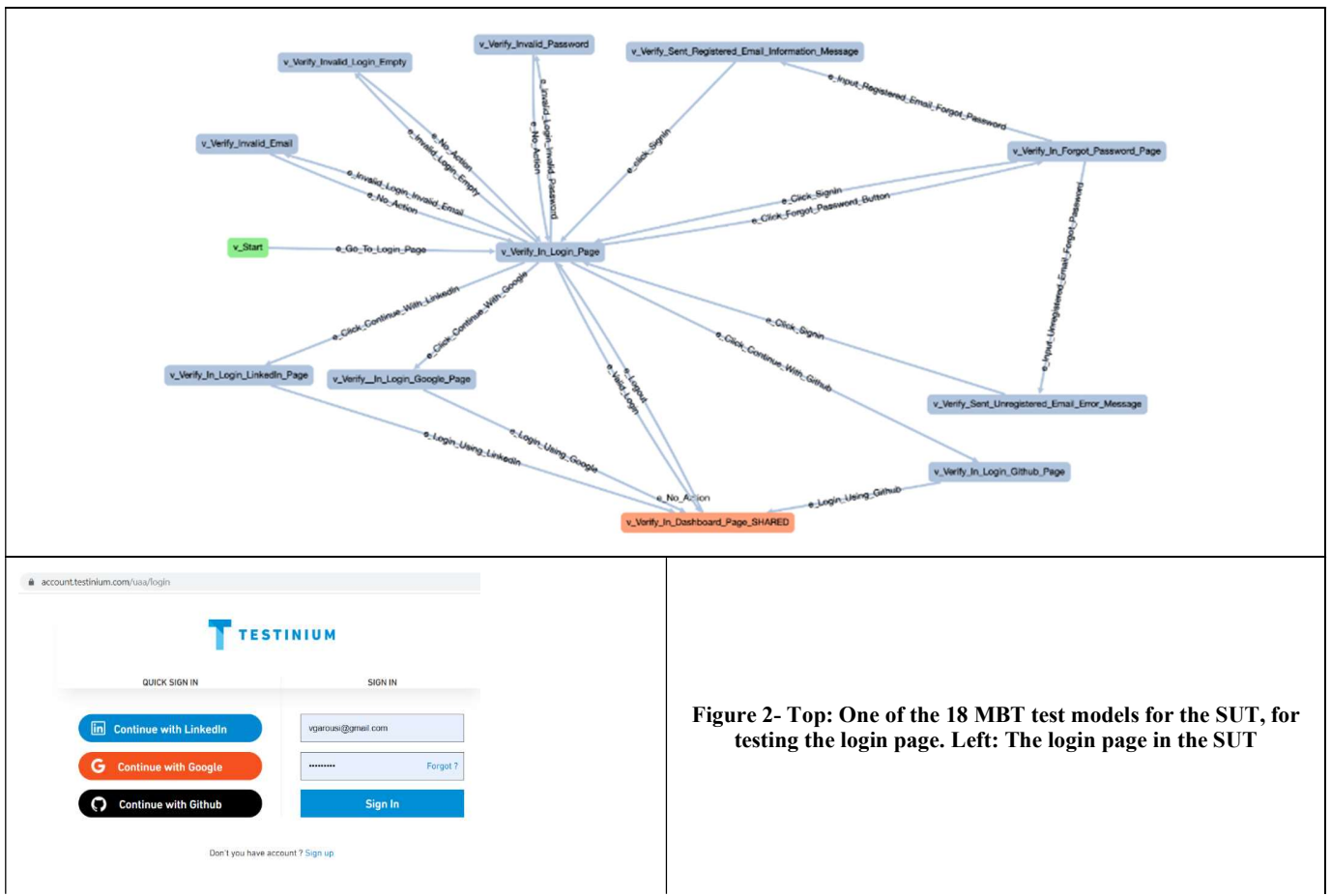

**Figure 2- Top: One of the 18 MBT test models for the SUT, for testing the login page. Left: The login page in the SUT**

To further showcase our MBT approach, and the usage of MBTCover tool, we have recorded several videos of the MBT test executions and the MBTCover in action and have posted them in a YouTube playlist (bit.ly/ VideosMBTTestinium).

### 3.3 Features of MBTCover

We show screenshots from MBTCover and its features in Figure 3. In these screenshots, the SUT (Figure 3b) is Testinium, and more precisely the *Dashboard* page (i.e., the main landing page) of the tool, which is shown to the user after successful login.

An in-progress execution of the MBT suite, using the GraphWalker tool, is shown in Figure 3c. The visible test model is the one testing the *Dashboard* page of the SUT.

MBTCover shows the coverage charts, which are updated every few seconds, a parameter which can be changed by the user. The chart in Figure 3a shows three coverage metrics in one line-chart view: front-end code-coverage ratio, back-end coverage, and requirement coverage.

We designed the server-side coverage line in the chart to be shown as cumulative values (Figure 3a), and thus the coverage values either increase or stay the same, as the time goes by. Once test execution finishes, or in the middle upon request of the user by pressing a button, MBTCover instructs the JaCoCo engine in the server side to export the server-side coverage report to HTML format and MBTCover provides a view of that detailed coverage report (Figure 3e).

In the current implementation of MBTCover, for the front-end, the tool presents two separate coverage charts (Figure 3d): (1) The top-view chart shows the *cumulative* front-end (JS) coverage, meaning that the coverage calculation has been done based on the combined lines of JS covered in all the web pages of the SUT, reached so far, divided by the sum of all JS code lines; (2) The bottom chart shows the JS coverage % of the *current* web page, being tested by the MBT suite so far. The decision to implement each of the above features in these specific manners have been as the result of discussions with several teams of test engineers in the company. For example, test engineers have provided the feedback to us that it would be useful for the purpose of testing to observe both cumulative and in-single-page coverage values while testing.

Our software development process for MBTCover has been iterative and Agile. During the development process, we have seen the need to also show, in the GUI of the tool, several important and useful statistics (Figure 3a), which include: (1) number of test models reached (covered) so far, (2) number of model nodes covered so far, and (3) number of nodes executed so far. Note that there is a difference between the last two mentioned items since the former is the node coverage of MBT models, while the latter is the number of nodes which have been executed and a given node could be visited more than once during test execution.

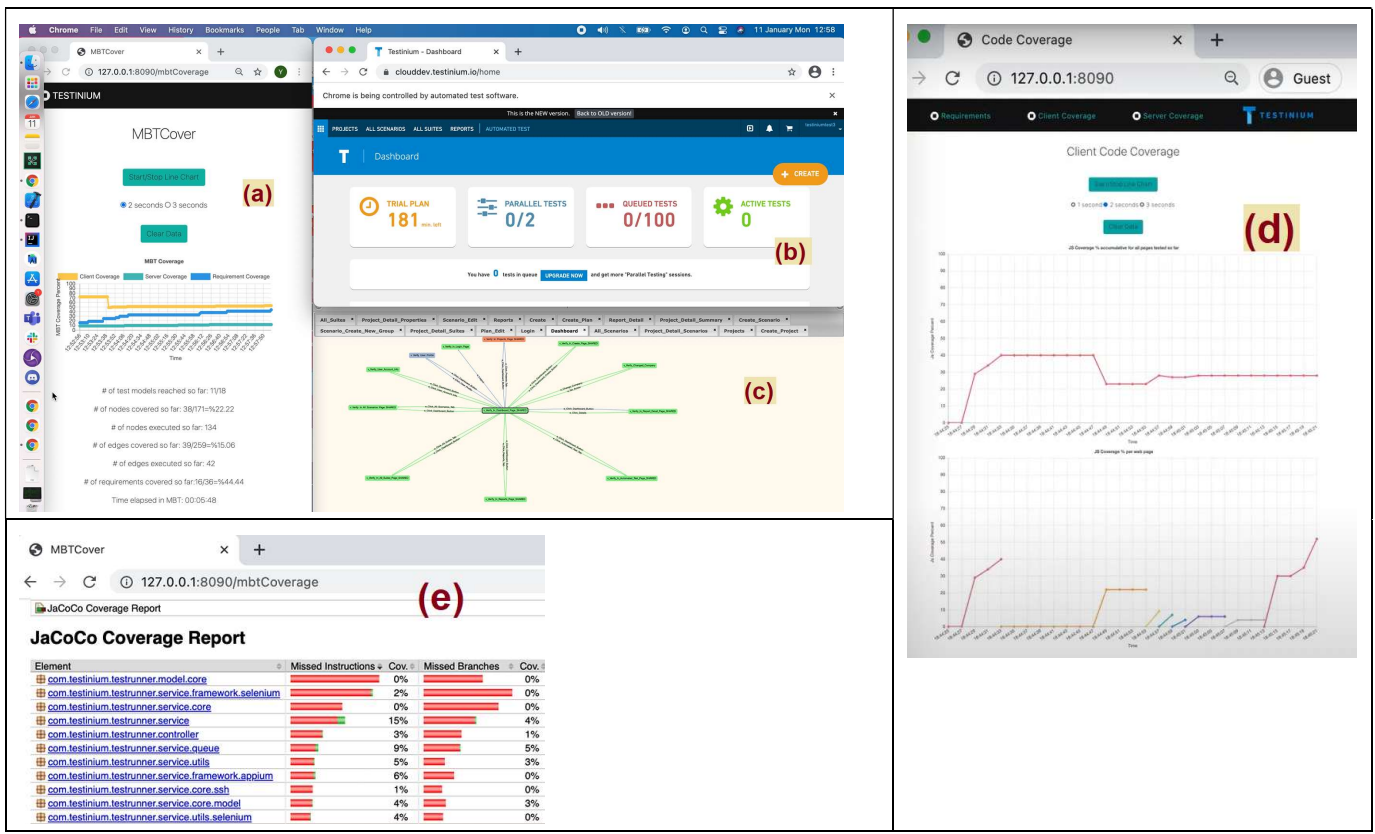

**Figure 3- (a, d and e): User interface of MBTCover -- (b): the SUT: Testinium -- (c): the MBT suite in execution**

Another important feature that our test engineers have mentioned the need for, in the tool, was live requirements coverage. The format of requirement documentation, practiced in the company, is use-cases. We thus defined requirements coverage as the ratio of the specified use-case steps covered so far by the MBT suite. To enable measurement of such a coverage, we specify the traceability between requirements and model elements (nodes or edges), using a feature of the GraphWalker tool, as shown in Figure 4.

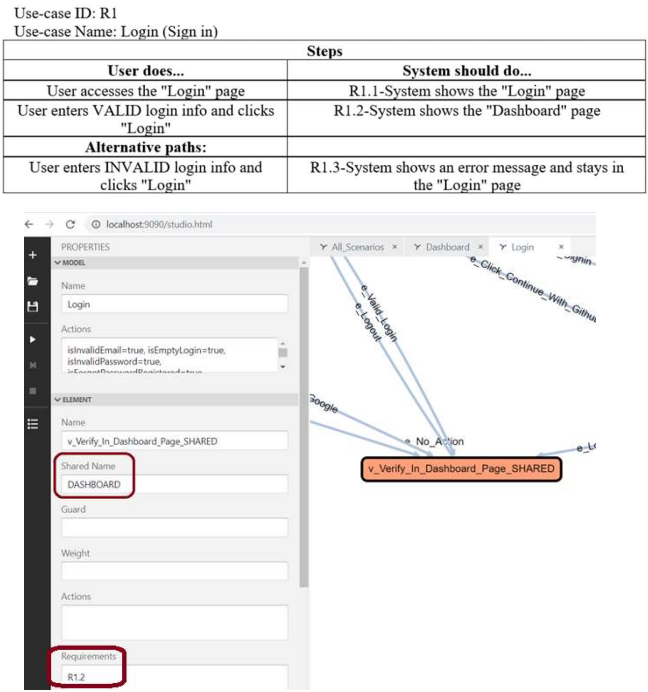

**Figure 4- Assigning a given node to a requirements item in the chosen MBT tool**

For this purpose, test engineers document the use-case descriptions and steps. For example, we show the use-case steps of the *Login* feature of the SUT (Testinium) in Figure 4.

GraphWalker allows specifying the requirements "tags", corresponding to a model element (a node or an edge). When a MBT suite is running using GraphWalker, we have programmed MBTCover to make calls to GraphWalker's API, and query the latest set of requirements covered, and use that information to calculate the percentage of requirements coverage (see the requirements coverage in Figure 3a).

The MBTCover is fully open-source in github.com/vgarousi/MBTCover. The online repository also includes the tool's design document, several SUT examples, their UML use-case requirements, and demo videos.

### 3.4 Interpreting the coverage trends in the above example

Let us briefly discuss the example coverage data in Figure 3, through which we also explain some of the many benefits of MBTCover. As seen in X-axis of Figure 3a, the coverage curves are from about 5-minute execution of the test models (minute 12:52 to 12:57).

In Figure 3d-top, we can see that the cumulative front-end (JS) coverage has increased from mid-0% to above 40% and then back to mid-20% as the MBT suite continues its execution and visits different web pages of the SUT. The reason for the fluctuation (up and downs) is that different web pages of the SUT use (reference) different JS files with different Line-of-Code (LOC) sizes. In other words, the JS coverage calculation, which is based on the formula: *num_of_covered_JS_lines / total_JS_lines*, would yield a lower value since the "divisor" value in the formula would increase suddenly when an additional JS file is imported during execution.

In Figure 3d-bottom, the JS coverage ratio of the current web page also provides valuable information, as we can see the extent of JS code coverage in the current page, as being tested by the MBT suite. For example, the MBT execution of Testinium starts with the *Login* page (Figure 2) and then moves to the *Dashboard* page (the first orange and the second yellow chart lines correspond to those two pages). As expected, the current web-page coverage chart resets to the value of 0% in each page and then *grows* up to a certain level, until the browser navigates to a different web page of the SUT, as the MBT suite is running.

Test engineers can see the extent of coverage in the current page and take actions if s/he decides to. For example, if the coverage is low, test engineers can investigate why most parts of the included JS files have not been covered; or they may decide to improve the MBT models, e.g., add more test paths. In terms of usage mode of the tool, together with our test engineers, we have experimented both "live" analysis of coverage data, and also have done video recording of the different coverage curves, for later analysis and replay by test engineers. Both modes have proved to be useful.

In Figure 3a, the server-side coverage value starts from about 10% as the MBT suite starts its execution and since the chart is a cumulative percentage, it will either stay constant or increase by time. The server-side coverage grows very slowly to low teens, and slowly changes in the five-minute snapshot of test execution. This is due to the extra-large size of Java application code-base on the server (the Testinium tool being the SUT), and the fact that MBT suite only covers a small ratio of that code-base during the 5-minute execution. Let us note that, for this particular MBT test suite, a full execution of the test suite with coverage takes about 6 hours on our high-end machines.

## 4 INDUSTRIAL EXPERIENCE REPORT: BENEFITS OF USING MBTCOVER IN TEST-AUTOMATION PROJECTS

After finalizing the stable release of MBTCover in 2022, it has been in active use in MBT projects of several internal tools of the company, e.g., Testinium and Loadium (a web-based load testing tool, loadium.io). Our test engineers are also actively using MBTCover in several (10+) test automation projects of our clients, e.g., the web and mobile apps of several airlines, and several financial and energy-related corporations.

In our conversations (semi-structured informal interviews) with eight test engineers in the company so far, their general consensus has been that MBTCover is a useful tool in exploring the scale and extent of MBT test automation, via the four consolidated coverage data that it provides: front-end, back-end, requirements and model coverage. While many of our testers have been using coverage for unit testing before this new tool, they reported that assessing coverage for GUI testing using a tool such as MBTCover is a niche and useful approach.

By using MBTCover, they have reported that they are able to: (1) pinpoint unused parts of JS files, in large test runs, and thus optimize the size of JS code files, (2) detect part of a given MBT test suite in not testing certain parts of front-end, back-end, and requirements, (3) improve MBT test suites to ensure full coverage w.r.t. any of the four coverage domains above (front-end, back-end, requirements and model coverage), (4) assess traceability of requirements to MBT models to code, which have been useful in various test regression and change-impact analysis scenarios.

## 5 DISCUSSIONS AND FUTURE WORKS

Our ongoing work in MBT since 2018 [2, 15, 16] has demonstrated the real-world need for industrial-scale coverage analysis tools, integrating different aspects of coverage for testing web and mobile applications. To address that need, we have designed, developed, open-sourced and have also evaluated the MBTCover tool. More than 10 test engineers are actively using the tool voluntarily in their daily tasks, and this shows the benefit of using the tool in MBT projects.

In this paper, we presented the features of MBTCover and our overall experience from using it in large test-automation projects. We are certain that other software test engineers, who use the MBT approach in their projects, could find the MBTCover tool useful in their contexts.

We plan to design and conduct empirical studies on the benefits of MBTCover, and to use stakeholders' feedback to further improve the tool in the future.